\documentclass[aps,twocolumn,nofootinbib,
preprintnumbers,superscriptaddress]{revtex4-1}
\usepackage{makecell}
\usepackage{latexsym}
\usepackage{siunitx}
\usepackage{graphics}
\usepackage{amsmath}
\usepackage{tikz}
\usepackage{amsfonts}
\usepackage{amsmath}
\usepackage{amssymb}
\usepackage[paper=letterpaper,margin=1in]{geometry}
\usepackage{braket}
\usepackage{upgreek}

\newcommand{\beq}{\begin{eqnarray}}
\newcommand{\eeq}{\end{eqnarray}}

\usepackage{tikz}

\newtheorem{thm}{Theorem}[section]

\newtheorem{preremark}[thm]{Remark}

\begin{document}
\title{ Stranger than Metals}

\author{Philip W. Phillips}
\affiliation{Department of Physics and Institute for Condensed Matter Theory,
University of Illinois, 1110 W. Green Street, Urbana, IL 61801}
\author{Nigel E. Hussey}
\affiliation{H. H. Wills Physics Laboratory, University of Bristol, Tyndall Avenue, Bristol BS8 1TL, United Kingdom}
\affiliation{High Field Magnet Laboratory (HFML-EMFL) and Institute for Molecules and Materials, Radboud University, Toernooiveld 7, 6525 ED Nijmegen, Netherlands}
\author{Peter Abbamonte}
\affiliation{Department of Physics,
University of Illinois, 1110 W. Green Street, Urbana, IL 61801}

\begin{abstract}
  Although the resistivity in traditional metals increases with temperature, its $T$ dependence vanishes at low or high temperature, albeit for different reasons. Here, we review a class of materials, known as \lq strange' metals, that can violate both principles. In materials exhibiting such behavior, the change in slope of the resistivity as the mean free path drops below the lattice constant, or as $T \rightarrow 0$, can be imperceptible, suggesting complete continuity between the charge carriers at low and high $T$.  Since particles cannot scatter at length scales shorter than the interatomic spacing, strange metallicity calls into question the relevance of locality and a particle picture of the underlying current.  This review focuses on transport and spectroscopic data on candidate strange metals with an eye to isolate and identify a unifying physical principle.  Special attention is paid to quantum criticality, Planckian dissipation, Mottness, and whether a new gauge principle, which has a clear experimental signature, is needed to account for the non-local transport seen in strange metals.  For the cuprates, strange metallicity is shown to track the superfluid density, thereby making a theory of this state the primary hurdle in solving the riddle of high-temperature superconductivity. 
\end{abstract}

\maketitle

To understand the essential tension between quantum mechanics and gravity, simply imagine two electrons impinging on the event horizon of a black hole.  While classical gravity predicts that they meet at the center, quantum mechanics forbids this should the electrons have the same spin.  In essence, classical gravity has no way of preserving Pauli exclusion. Of course replacing classical general relativity with a quantum theory of gravity at small enough scales
resolves the problem,
but what is this scale? In 1899, Planck formulated a universal length now regarded as the scale below which a quantum theory of gravity supplants its classical counterpart.  The Planck scale, 
\beq
\ell_P=\sqrt{\frac{\hbar G}{c^3}},
\eeq
is pure dimensional analysis on three fundamental constants: the speed of light, $c$, Newton's gravitational constant, $G$, and the quantum of uncertainty, $\hbar$, Planck's constant, $h$, divided by $2\pi$. This leads naturally to a Planck time as the ratio of the Planck length to the speed of light, $\ell_P/c$.  Such a Planckian analysis can be extended equally to many-body systems in contact with a heat bath.  All that is necessary is to include the temperature $T$. A similar dimensional analysis then leads to  
\beq
\label{plank}
\tau_P=\frac{\hbar}{k_BT}
\eeq
as the shortest time for heat loss in a many-body system obeying quantum mechanics with $k_B$, Boltzmann's constant. As no system parameters enter $\tau_P$, this quantity occupies a similar fundamental role in analogy with the Planck length and is referred to as the Planckian dissipation time. Although Eq. (\ref{plank}) has had previous incarnations \cite{matsubara,chn89}, in the realm of charge transport, it defines the time scale for scale-invariant or Planckian dissipation \cite{zaanen04}.   Scale-invariance follows because there is no scale other than temperature appearing in $\tau_P$.  Achieving such scale invariance necessitates a highly entangled many-body state. Such a state would lead to a breakdown of a local single-particle and the advent of new collective non-local entities as the charge carriers.  Precisely what the new propagating degrees of freedom are is the key mystery of the strange metal.

\begin{figure}
\centering
\includegraphics[scale=0.45]{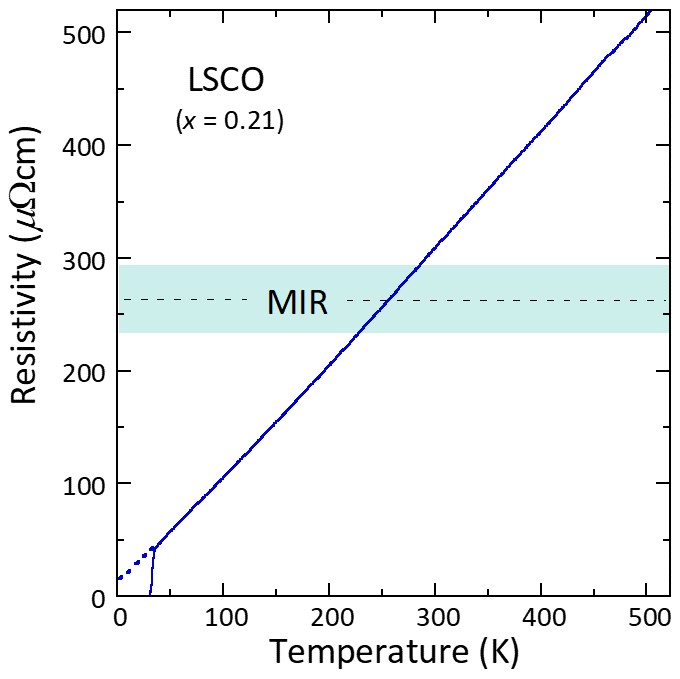}
\caption{
In-plane resistivity of La$_{2-x}$Sr$_x$CuO$_4$ ($x$ = 0.21, adapted from Ref.~\cite{cooper09, hussey11}). The dotted points are extrapolated from high-field magnetoresistance data  \cite{cooper09}. The shaded area shows the Mott-Ioffe-Regel (MIR) boundary where the mean-free-path becomes comparable to the interparticle scattering length.}
\label{lsco}
\end{figure}

While the Planck scale $\ell_P$ requires high-energy accelerators much beyond anything now in use, such is not the case with physics at the Planckian dissipation limit. Early table-top experiments on cuprate superconductors, for example, revealed a \lq strange metal' regime defined by a robust $T$-linear resistivity extending to the highest temperatures measured \cite{gurvitch87, martin90, takagi92} (see Fig. \ref{lsco}), a possible harbinger of Planckian dissipation.   Recall that in a Fermi liquid, the conductivity, can be well described by a Drude formula,
\beq
\sigma=\frac{n_e e^2}{m}\tau_{\rm tr}
\eeq
where $n_e$ is the charge carrier density, $e$ and $m$ the charge and mass of an electron, respectively, and the transport lifetime 
\beq
\tau_{\rm tr}=\frac{\hbar E_F}{(k_BT)^2}=\frac{E_F}{k_BT}\tau_P,
\eeq
contains the Fermi energy $E_F$ of the quasiparticles. No such energy scale appears in Eq. (\ref{plank}). If the scattering rate in cuprates is directly proportional to the resistivity, as it is in simple metals, $T$-linear resistivity is equivalent to scale-invariant Planckian dissipation only if $\tau_{tr}=\alpha_1 \tau_P$ with $\alpha_1\sim 1$.  While this state of affairs seems to be realized in a host of correlated metals, including the cuprates \cite{marel03,cooper09,legros19,bruin13}, questions that deserve further deliberation are how accurately is $\alpha_1$ known and what are the assumptions that go into its determination?  Regardless of the possible relationship with Planckian dissipation, what makes $T$-linear resistivity in the cuprates truly novel is its persistence -- from mK temperatures (in both the electron- and hole-doped cuprates) \cite{fournier98, mackenzie96b} up to 1000 K (in the hole-doped cuprates) \cite{gurvitch87,takagi92} -- and its omnipresence, the strange metal regime dominating large swathes of the temperature vs.~doping phase diagram \cite{nagaosa92}. In normal metals \cite{iofferegel,gurvitch81} as well as some heavy fermions \cite{husseyMIR}, the resistivity asymptotically approaches a saturation value commensurate with the mean-free-path $\ell$  becoming comparable with the interatomic spacing $a$ -- the minimum length over which a Bloch wave and its associated Fermi velocity and wave vector can be defined. In many correlated metals -- collectively refered to as \lq bad metals' -- $\ell<a$ at high $T$, thereby violating the so-called Mott-Ioffe-Regel (MIR) limit \cite{iofferegel, mott, husseyMIR, martin90, takagi92, hussey11}. Remarkably, no saturation occurs in these bad metals across the MIR threshold, implying that the whole notion of a Fermi velocity of quasiparticles breaks down at high $T$. In certain cases, an example of which is shown in Fig.~\ref{lsco}, there is no discernible change in slope as the MIR limit is exceeded. While this circumstance occurs only in a narrow doping window (in cuprates) \cite{hussey11}, such continuity does suggest that, even at low $T$, quasiparticles \cite{emkiv95} cannot be the effective propagating degrees of freedom. Evidently, in strongly correlated electron matter, the current-carrying degrees of freedom in the IR need not have a particle interpretation. Precisely what the charge carriers are and the experimental delineation of the strange metal will be the subject of this review. 

Over time, the label \lq strange metal' has seemingly become ubiquitous, used to describe any metallic system whose transport properties display behavior that is irreconcilable with conventional Fermi-liquid or Boltzmann transport theory. This catch-all phraseology, however, is unhelpful as it fails to differentiate between the various types of non-Fermi-liquid behavior observed, some of which deserve special deliberation on their own. In this review, we attempt to bring strange metal phenomenology into sharper focus, by addressing a number of pertinent questions. Does the term refer to the resistive behavior of correlated electron systems at high or low temperatures or both? Does it describe {\it any} $T$-linear resistivity associated with the Planckian timescale, or something unique? Does it describe the physics of a doped Mott insulator or the physics associated with quantum criticality (whose underlying origins may or may not include Mottness as a key ingredient)? Finally, does anything local carry the current and if not, does explicating the propagating degrees of freedom in the strange metal require a theory as novel as quantum gravity?

\section{Is Strange Metallicity Ubiquitous?}

\begin{table*}
\begin{ruledtabular}
\caption{Summary of the dc transport properties of various strange metal candidates. The first column identifies the candidate compound or family of compounds. For the hole-doped cuprates, underdoped (UD), optimally doped (OP) and overdoped (OD) compounds are treated separately, but individual compounds within each sub-set are not listed as their transport properties are found to be generic. For the electron-doped cuprates, only La$_{2-x}$Ce$_x$CuO$_4$ is selected since this is the material for which all relevant properties have been studied, though the Pr- and Nd-based sister compounds do show similar behavior. \lq MATBG' stands for magic-angle twisted bilayer graphene. The second column considers bad metallic behavior, though here a $\checkmark$ mark refers only to those materials that exhibit $T$-linear resistivity beyond the Mott-Ioffe-Regel (MIR) limit. Systems identified with a $\times$ show either a tendency towards saturation or a marked reduction in slope near the MIR limit. $\checkmark$ marks in the third column identify systems that at a singular point in their respective phase diagram(s), exhibit $T$-linear resistivity down to the lowest temperatures studied thus far. The $\lq$ extended criticality' heading for column 4 refers then to systems where a predominant $T$-linear resistivity at low-$T$ extends over a finite region of the phase diagram. Column 5 considers systems that exhibit a $T^2$ dependence of the inverse Hall angle cot$\Theta_{\rm H}$ in the same temperature range where $\rho(T)$ is $T$-linear. Compounds satifsying the \lq Modified Kohler's' label in column 6 have a low-field magnetoresistance (MR), defined as $\rho(H,T)-\rho(0,T)/\rho(0,T)$, that exhibits a similar $T$-dependence to tan$^2\Theta_{\rm H}$. The last two columns inspect the high-field MR behavior of strange metal candidates. Note that the observation of a $H$-linear MR at high fields does not imply the form of the MR over all fields and temperatures exhibits quadrature scaling. La$_{2-x}$Sr$_x$CuO$_4$, for example, displays simultaneous $H$- and $T$-linearity but no quadrature scaling. The * marks for FeSe$_{1-x}$S$_x$ highlight the fact that the $H$-linear/quadrature MR seen in this family coexists with a more conventional MR contribution, indicating the presence of both strange metal and Fermi-liquid-like components in the dc transport. The ** marks alongside YbAlB$_4$ highlight the fact that while $T$-linear resistivity is observed over a wide pressure range, its limiting low-$T$ dependence is $T^{1.5}$. Finally, dash marks indicate where, as yet, there have been no reports confirming or otherwise the considered behavior.\\}
\begin{tabular}{cccccccc}
\centering
 & \makecell{$\rho\propto T$} & \makecell{$\rho\propto T$} & \makecell{Extended} & \makecell{cot$\Theta_{\rm H}\propto T^2$} & \makecell{Modified Kohler's} & \makecell{$H$-linear MR} & \makecell{Quadrature} \\ & \makecell{as $T$$\rightarrow \infty$} & \makecell{as $T$$\rightarrow$ 0} & \makecell{criticality} & \makecell{(at low $H$)} & \makecell{(at low $H$)} & \makecell{(at high $H$)} & \makecell{MR}\\
\colrule
UD $p$-cuprates & $\checkmark$ \cite{takagi92} & $\times$ \cite{proust16} & $\times$ \cite{barisic13} & $\checkmark$ \cite{carrington92} & $\checkmark$ \cite{chan14} & - & - \\
OP $p$-cuprates & $\checkmark$ \cite{gurvitch87} & - & - & $\checkmark$ \cite{chien91} & $\checkmark$ \cite{harris95} & $\checkmark$ \cite{giraldo18} & $\times$ \cite{boyd19} \\
OD $p$-cuprates & $\checkmark$ \cite{takagi92} & $\checkmark$ \cite{cooper09} & $\checkmark$ \cite{cooper09} & $\checkmark$ \cite{manako92} & $\times$ \cite{ayres20} & $\checkmark$ \cite{ayres20} & $\checkmark$ \cite{ayres20} \\
La$_{2-x}$Ce$_x$CuO$_4$ & $\times$ \cite{poniatowski20} & $\checkmark$ \cite{jin11} & $\checkmark$ \cite{jin11} & $\times$ \cite{li07} & $\times$ \cite{poniatowski21} & $\checkmark$ \cite{sarkar18} & $\times$ \cite{sarkar18} \\
Sr$_2$RuO$_4$ & $\checkmark$ \cite{tyler98} & $\times$ \cite{hussey98} & $\times$ \cite{barber18} & $\times$ \cite{mackenzie96} & $\times$ \cite{hussey98} & $\times$ \cite{hussey98} & $\times$ \cite{hussey98}\\
Sr$_3$Ru$_2$O$_7$ & $\checkmark$ \cite{bruin13} & $\checkmark$ \cite{bruin13} & $\times$ \cite{bruin13} & $\times$ & - & - & - \\
FeSe$_{1-x}$S$_x$ & $\times$ \cite{kasahara14} & $\checkmark$ \cite{licci19a} & $\times$ \cite{licci19a} & $\checkmark$ \cite{huang20} & $\checkmark$ \cite{huang20} & $\checkmark$* \cite{licci19b} & $\checkmark$* \cite{licci19b} \\
BaFe$_2$(As$_{1-x}$P$_x$)$_2$ & $\times$ \cite{hu18} & $\checkmark$ \cite{analytis14} & $\times$ \cite{analytis14} & - & $\checkmark$ \cite{kasahara10} & $\checkmark$ \cite{hayes16} & $\checkmark$ \cite{hayes16} \\
Ba(Fe$_{1/3}$Co$_{1/3}$Ni$_{1/3}$)$_2$As$_2$ & - & $\checkmark$ \cite{nakajima20} & $\times$ \cite{nakajima20} & - & - & $\checkmark$ \cite{nakajima20} & $\checkmark$ \cite{nakajima20} \\
YbRh$_2$Si$_2$ & $\times$ \cite{trovarelli00} & $\checkmark$ \cite{custers03} & $\checkmark$ \cite{custers10} & $\checkmark$ \cite{paschen04} & - & - & - \\
YbBAl$_4$ & $\times$ \cite{tomita15} & $\checkmark^{**}$ \cite{tomita15} & $\checkmark^{**}$ \cite{tomita15} & - & - & - & - \\
CeCoIn$_5$ & $\times$ \cite{nakajima07} & $\checkmark$ \cite{bianchi03} & $\times$ \cite{bianchi03} & $\checkmark$ \cite{nakajima07} & $\checkmark$ \cite{nakajima07} & - & - \\
CeRh$_6$Ge$_4$ & $\times$ \cite{shen20} & $\checkmark$ \cite{shen20} & $\times$ \cite{shen20} & - & - & - & - \\
(TMTSF)$_2$PF$_6$ & - & $\checkmark$ \cite{doiron09} & $\checkmark$ \cite{doiron09} & - & - & - & - \\
MATBG & $\checkmark$ \cite{polshyn19} & $\checkmark$ \cite{cao20} & $\checkmark$ \cite{cao20} & $\checkmark$ \cite{lyu20} & - & - & -
\end{tabular}
\label{tab:SMsurvey}
\end{ruledtabular}
\end{table*}

In addressing this question, we must first acknowledge the many definitions of strange metallic behavior that exist, the simplest being a material hosting a metallic-like resistivity in the absence of quasiparticles. A more precise, if empirical, definition centres on the $T$-linear resistivity, specifically one that is distinguishable from that manifest in simple metals and attributed to electron-phonon scattering. For a metal to be classified as strange, the $T$-linearity  must extend far beyond the typical bounds associated with phonon-mediated resistivity. At low $T$, this is typically one third of the Debye temperature, while at high $T$, it is once the magnitude of the resistivity approaches (roughly 1/2) the value commensurate with the MIR limit. A sub-set of correlated metals, such as SrRuO$_3$ \cite{allen96} and Sr$_2$RuO$_4$ \cite{tyler98}, exhibit $T$-linear resistivity at high-$T$ with a magnitude that clearly violates the MIR limit, but as the system cools down, conventional Fermi-liquid behavior is restored \cite{mackenzie98, hussey98}. Hence, while they are {\it bona fide} bad metals -- exhibiting metallic resistivity beyond the MIR limit --  they do not classify as strange \cite {gunnarsson03, husseyMIR}. 

Another subset, identified here as quantum critical metals, exhibit $T$-linear resistivity down to the lowest temperatures studied, but only at a singular quantum critical point (QCP) in their phase diagram associated with a continuous quantum phase transition to a symmetry broken phase that occurs at $T$ = 0. In most cases, the phase transition in question is associated with finite-\textbf{Q} antiferromagnetism (as in pure YbRh$_2$Si$_2$ \cite{trovarelli00}, CeCoIn$_5$ \cite{bianchi03} and BaFe$_2$(As$_{1-x}$P$_x$)$_2$ \cite{analytis14}) though recently, similar behavior has also been reported in systems exhibiting zero-\textbf{Q} order, such as nematic FeSe$_{1-x}$S$_x$ \cite{licci19a} or ferromagnetic CeRh$_6$Ge$_4$ \cite{shen20}. Away from the QCP, the low-$T$ resistivity recovers the canonical $T^2$ Fermi-liquid form, albeit with a coefficient that is enhanced as the QCP is approached and the order parameter fluctuations soften.

By contrast, in overdoped cuprates (both hole- \cite{cooper09, legros19} and electron-doped \cite{jin11}), Ge-doped YbRh$_2$Si$_2$ \cite{custers10}, YbBAl$_4$ \cite{tomita15} and the organic Bechgaard salts \cite{doiron09}, $\rho(T)$ is predominantly $T$-linear down to low-$T$ not at a singular point in their respective phase diagrams but over an extended range of the relevant tuning parameter. At first sight, this \lq extended criticality' is difficult to reconcile with current theories of quantum criticality, which predict a crossover to a purely $T^2$ resistivity and thus a recovery of FL behavior at low $T$ everywhere except at the (singular) QCP. Arguably, it is this feature -- incompatibility with standard Fermi-liquid {\it and} quantum critical scenarios -- that distinguishes a geniune strange metal from its aspirants. Intriguingly, in many of these systems $\alpha_1$ -- the coefficient of the $T$-linear resistivity -- is found to scale with the superconducting transition temperature $T_c$. Moreover, for La$_{2-x}$Ce$_x$CuO$_4$ \cite{jin11} and (TMTSF)$_2$PF$_6$ \cite{doiron09}, extended criticality emerges beyond a spin density wave QCP, suggesting an intimate link between the strange metal transport, superconductivity and the presence of critical or long-wavelength spin fluctuations. In hole-doped cuprates, however, the strange metal regime looks different, in the sense that the extended criticality emerges beyond the end of the pseudogap regime that does not coincide with a magnetic quantum phase transition \cite{hussey18}. Furthermore, while the pseudogap plays host to a multitude of broken symmetry states, the jury is still out as to whether any of these are responsible for pseudogap formation or merely instabilities of it. 

Besides $T$-linear resistivity, strange metals also exhibit anomalous behavior in their magnetotransport, including 1) a quadratic temperature dependence of the inverse Hall angle $\cot\Theta_H=\sigma_{\rm xy}/\sigma_{\rm xx}$,  2) a transverse magnetoresistance (MR) that at low field exhibits modified Kohler's scaling ($\Delta\rho/\rho(0) \propto$ tan$^2\Theta_{\rm H} \propto (1/T^2)^2$ or $(1/(A+BT^2)^2)$ \cite{harris95}) and/or 3) a $H$-linear MR at high fields that may or may not follow quadrature scaling (whereby $\Delta\rho/T \propto$ $\sqrt{1+\gamma(H/T)^2}$) \cite{hayes16,licci19b}. A survey of the dc transport properties of several strange metal candidates is presented in Table \ref{tab:SMsurvey}. The combination of a modified Kohler's rule and $T^2$ Hall angle has been interpreted to indicate the presence of distinct relaxation times, either for different loci in momentum space \cite{carrington92} or for relaxation processes normal and tangential to the underlying Fermi surface \cite{chien91}. The $H$-linear MR, on the other hand, is inextricably tied to the $T$-linear zero-field resistivity via its $H/T$ scaling relation, a relation that can also extend over a broad range of the relevant tuning parameter \cite{ayres20}. In some cases, this link can be obscured, either because $\rho(T)$ itself is not strictly $T$-linear \cite{ayres20} or because the quadrature MR co-exists with a more conventional orbital MR \cite{licci19b}. Both sets of behavior highlight once again the possible coexistence of two relaxation times or two distinct charge-carrying sectors in real materials. Curiously, quadrature scaling does breaks down inside the pseudogap regime \cite{giraldo18, boyd19} while modified Kohler's scaling is recovered \cite{harris95, chan14}, suggesting that the two phenomena may be mutually exclusive in single-band materials. In multiband materials such as FeSe$_{1-x}$S$_x$, on the other hand, these different manifestations of strange metallic transport appear side-by-side \cite{licci19b,huang20}. Irrespective of these caveats and complexities, what is striking about the quadrature MR is that it occurs in systems with distinct Fermi surface topologies, dominant interactions and energy scales, hinting at some universal, but as yet unidentified, organizing principle.

Restricting the strange metal moniker, as done here, to materials that exhibit low-$T$ $T$-linear resistivity over an extended region of phase space likewise restricts strange metallicity to a select \lq club'. What shared feature binds them together is the key question that will be explored in the coming sections.

\section{Is it Quantum Critical?}

Such scale-free $T$-linear resistivity is highly suggestive of some form of underlying quantum criticality in which the only relevant scale is the temperature governing collisions between excitations of the order parameter \cite{damlesachdev97}. In fact, following the advent of marginal Fermi liquid (MFL) phenomenology with its particular charge and spin fluctuation spectra and associated ($T, \omega$)-linear self energies \cite{varma96}, the common interpretation of such $T$-linear resistivity was and still remains the nucleus of ideas centered on quantum criticality. The strict definition of quantum criticality requires the divergence of a thermodynamic quantity. In heavy fermion metals, the electronic heat capacity ratio $C_{\rm el}/T$ indeed grows as $\ln (1/T)$ as the antiferromagnetic correlations diverge \cite{hf1,hf2,bianchi03}. In certain hole-doped cuprates, $C_{\rm el}/T$ also scales as $\ln (1/T)$ at doping levels close to the end of the pseudogap regime \cite{MichonSheat} though here, evidence for a divergent length scale of an associated order parameter is currently lacking \cite{tallon}.  Moreover, photoemission suggests that at a $T$-independent critical doping $p_c\approx 0.19$, all signatures of incoherent spectral features that define the strange metal cease, giving way to a more conventional coherent response \cite{chen19}. The abruptness of the transition suggests that it is first-order, posing a challenge to interpretations based solely on criticality.

As touched upon in the previous section, another major hurdle for the standard criticality scenario is that the $T$-linear resistivity persists over a wide range of the relevant tuneable parameter, be it doping as is the case for cuprates \cite{cooper09,jin11,hussey13,legros19} and MATBG \cite{cao20}, pressure for YbBAl$_4$ \cite{tomita15} and the organics \cite{doiron09} or magnetic field for Ge-doped YbRh$_2$Si$_2$ \cite{custers10}. If quantum criticality is the cause, then it is difficult to fathom how a thermodynamic quantity can be fashioned to diverge over an entire phase.

\begin{figure}
\centering
\includegraphics[scale=0.5]{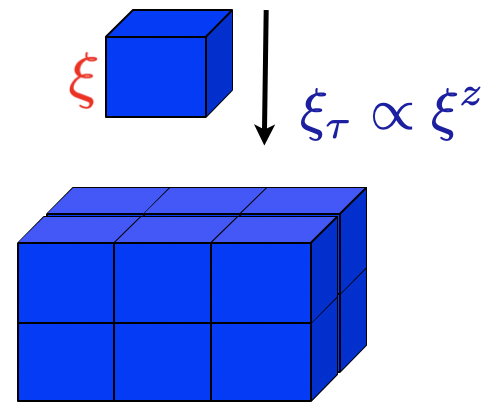}
\caption{Single-parameter scaling hypothesis in which all length and time scales are governed by the same diverging length scale, $\xi$ which diverges at the quantum critical point as $(g-g_c)^{-\nu}$, $g$ the coupling constant for the critical interaction. $z$ is the dynamical critical exponent.
}
\label{scaling}
\end{figure}

Despite these difficulties, it is worth exploring the connection $T$-linear resistivity has with continuous quantum critical phenomena, which for the sake of argument we presume to be tied to a singular point in the phase diagram. Regardless of the origin of the QCP, universality allows us to answer a simple question:  What constraints does quantum criticality place on the $T$-dependence of the resistivity? The answer to this question should just be governed by the fundamental length scale for the correlations.  The simplest formulation of quantum criticality is single-parameter scaling in which the spatial and temporal correlations are governed by the same diverging length (see Fig.~(\ref{scaling})). Making the additional assumption that the relevant charge carriers are formed from the quantum critical fluctuations, a simple scaling analysis on the singular part of the free energy results in the scaling law \cite{chamon05}
\beq
\label{qctemp}
\sigma(\omega=0,T)=\frac{q^2}{\hbar} f(\omega=0)\left(\frac{k_BT}{\hbar c}\right)^{(d-2)/z}
\eeq
for the $T$-dependence of the conductivity where $f(\omega=0)$ is a non-zero constant, $q$ is the charge and $z$ is the dynamical exponent, which from causality must obey the inequality $z\ge 1$. Absent from this expression is any dependence on an ancillary energy scale for example $E_F$ or the plasma frequency $\omega_p$ as the only assumption is scale-invariant transport irrespective of the details of the system. The analogous expression for the optical conductivity is \cite{wen92} 
\beq
\sigma(\omega,T=0)\propto \omega^{(d-2)/z}.
\label{qcomega}
\eeq  
In pure YbRh$_2$Si$_2$, for example, $\sigma^{-1}(\omega)$ follows an $\omega$-linear dependence at low frequencies in the same region of the ($T,H$) phase diagram -- the quantum critical \lq fan' --  where $\rho(T)$ is also linear, consistent with this notion of single-parameter scaling \cite{prochaska}. In cuprates, on the other hand, the situation is more nuanced. At intermediate frequencies -- sometimes referred to as the mid-infrared response -- $\sigma(\omega$) exhibits a ubiquitous $\omega^{-2/3}$ dependence \cite{marel03}. While this feature in $\sigma(\omega)$ has been interpreted in terms of quantum critical scaling \cite{marel03}, it is inconsistent with the single-parameter scaling described above. At any doping level, $\sigma(\omega)$ in the cuprates exhibits a minimum at roughly the charge transfer scale of 1 eV. This is traditionally \cite{marelcolorchange, CooperUVIR} used as the energy scale demarcating the separation between intraband and interband transitions and hence serves to separate the low-energy from the high-energy continua. It has long been debated whether the broad sub-eV $\sigma(\omega)$ response in cuprates is best analysed in terms of one or two components \cite{tannerdrude,CooperUVIR}. In the former, the $\omega^{-2/3}$ tail is simply a consequence of the strong $\omega$-linear dependence in 1/$\tau_{tr}(\omega)$ -- \`a la MFL -- while in the latter, it forms part of an incoherent response that is distinct from the coherent Drude weight centred at $\omega=0$ which itself is described with either a constant or $\omega$-dependent scattering rate.

Returning to the dc resistivity, we find that in cuprates, where $d$ = 3, an exponent $z=-1$ is required, a value that is strictly forbidden by causality \cite{chamon05}. For $d$ = 2, as in the case of MATBG, the $T$-dependence vanishes. This is of course fixed with the replacement of $d\rightarrow d^\ast=1$ for both materials. While $d^\ast$ can be construed as the number of dimensions \cite{shl} transverse to the Fermi surface, it is difficult to justify such a procedure here as the persistence of $T$-linearity with no change in slope above and below the MIR requires a theory that does not rely on FL concepts such as a Fermi velocity or energy. Furthermore, it is well known that introducing $d^\ast$ yields a power law for the heat capacity, $C\propto T^{3/2}$ which is not seen experimentally \cite{loramSH}.  On dimensional grounds, the $z=-1$ result in the context of the Drude formula is a consequence of compensating the square power of the plasma frequency with powers of $T$ so that the scaling form Eq. (\ref{qctemp}) is maintained. A distinct possibility is that perhaps some other form of quantum criticality beyond single-parameter scaling, such as a non-critical form of the entropy suggested recently \cite{zaanenentropy}, is at work here. We shall return to this idea in section V.

Another critical feature of the conductivity is its behavior at finite wave vector $k$ which may be quantified by the dynamic charge susceptibility,
\begin{equation}
\chi^{\prime\prime}(k,\omega)=-\frac{k^2}{\omega e^2} \Re\sigma(k,\omega),
\end{equation}
determined from electron energy-loss spectroscopy (EELS). A restriction on EELS is that it measures the longitudinal charge response while optics yields the transverse.  At vanishing momentum both are expected to be equal. As optics has no momentum resolution, comparison with EELS can only be made as $k\rightarrow 0$. The primary charge excitation in strange metals is a 1 eV plasmon that was long believed to exhibit the same behavior as in a normal Fermi liquid \cite{nucker1989,nucker1991}. Recent high-resolution M-EELS measurements have called this belief into question, showing that the large-$k$ response is dominated by a continuum  which remains flat to high energies, roughly 2 eV \cite{vig17,mitrano18,husain19}.  Such behavior is reminiscent of the MFL \cite{varma96} scenario except in that picture, the continuum persists up to a cut-off scale determined by the temperature not the Mott scale of 2 eV.  In addition, the continuum exhibits scale-invariant features but with a dynamical critical exponent, $z \sim \infty$, not possible from a simple QCP. 

We conclude then that no form of traditional quantum criticality can account easily for the power laws seen in strange metallic transport (though we recognize that $T$-linear resistivity {\it is} observed above what appear to be genuine singular QCPs). The photoemission experiments \cite{chen19} indicating a first-order transition pose an additional problem exacerbated by the possibility that the criticality might be relevant to a whole region \cite{cooper09, legros19,greene,dessau,cao20,tomita15,doiron09,custers10,hussey18} rather than a point. Such criticality over an extended region is reminiscent of critical charged matter \cite{kiritsis2,kiritsis1} arising from dilatonic models in gauge-gravity duality. We will revisit aspects of these ideas in a later section as they have been the most successful (see Table \ref{SMtheory}) thus far in reproducing the various characteristics of strange metal physics.

\section{Is it Planckian?}
 
While the electrical resistivity in metals can be measured directly, the scattering rate is entirely an inferred quantity. Herein lies the catch with Planckian dissipation.
Angle-resolved photoemission (ARPES) experiments on cuprates as early as 1999 reported that the width of the momentum distribution curves (MDCs) at optimal doping along the nodal direction ($(0,0)$ to $(\pi,\pi)$) scale as a linear function of temperature and $a_0$ + 0.75$\omega$ for frequencies that exceed $2.5k_BT$ \cite{valla99}. The momentum linewidth, which in photoemission enters as Im$\Sigma$ -- the imaginary part of the self energy -- can be used to define a lifetime through
\beq
\hbar v_k\Delta k={\rm Im}\Sigma(\bf k,\omega)=2\frac{\hbar}{\tau},
\eeq
with $v_k$ the group velocity for momentum state $k$.  Extracting the slope from the data in Figure (2) of Ref.~\cite{valla99} and using the experimentally reported Fermi velocity $v_F$ = 1.1 eV/\SI{}{\angstrom}, we find that the single-particle scattering rate $\hbar/\tau \sim 1.7 k_B T$, i.e. of order the Planckian limit. Similar results were obtained in subsequent ARPES studies \cite{kaminski05,bok10,dama03} with a key extension added by Reber {\it et al.} \cite{dessau} whereby the width of nodal states was observed to obey the quadrature form indicative of a power-law liquid, $((\hbar\omega)^2+(\beta k_BT)^2)^\lambda$ where $\lambda$ is a doping-dependent parameter equal to $1/2$ at optimal doping. 

This extraction of the scattering rate from ARPES, however, is not entirely problem-free as $v_F$ is hard to define in ARPES experiments at energies close to the Fermi level and where, for the most part, the width of the state exceeds its energy. Indeed, the integral of the density of states using as input the $v_F$ extracted from APRES measurements is found to account for only half of the as-measured electronic specific heat coefficient \cite{yoshida07}. Furthermore, this reliance on Fermiology leaves open the precise meaning of Fig. (2) of Ref.~\cite{bruin13} in which $\tau$ is plotted versus $v_F$ for a series of materials that violate the MIR limit at intermediate to high temperatures. Despite this, a similar extraction by Legros and colleagues \cite{legros19}, again using Fermiology but focusing on the low-$T$ resistivity, also found a transport scattering rate close to the Planckian bound. This consistency between the two analyses reflects the curious fact that the $T$-linear slope of the dc resistivity does not vary markedly as the MIR threshold is crossed. It does not, however, necessarily justify either approach in validating $T$-linear scattering at the Planckian limit. Finally, while $T$-linearity and Planckian dissipation appear synonymous in the cuprates, this is not universally the case. In YbRh$_2$Si$_2$ \cite{prochaska}, for example, the $T$-linear scattering rate is found to deviate strongly from the Planckian limit with $\tau_{tr} \sim 0.1\tau_P$ \cite{paschen04}, while in the electron-doped cuprates, the notion of a Planckian limit to the scattering rate has recently been challenged \cite{poniatowski21b}. This certainly adds to the intrigue regarding quantum criticality as the underlying cause of Planckian dissipation.

In principle, the optical conductivity permits an extraction of $\tau$ without recourse to Fermiology. Within a Drude model, the optical conductivity,
\beq
\sigma(\omega)=\frac{1}{4\pi}\frac{\omega_p^2\tau_{\rm tr}}{1+i\omega\tau_{\rm tr}},
\eeq
contains only $\tau_{\rm tr}$ and $\omega_p=\sqrt{4\pi n_e e^2/m}$. At zero frequency, the Drude formula naturally yields the dc conductivity $\sigma_{\rm dc}$ while an estimate for the relaxation rate can be extracted from the width at half maximum of the full Drude response. However, there is an important caveat: $\tau_{\rm tr}$ is frequency dependent in the cuprates, a condition that is consistent with various physical models including both the Fermi liquid and MFL scenarios as well as the large body \cite{dessau,valla99} of MDC analysis performed on the cuprates. While this prevents a clean separation of the conductivity into coherent and incoherent parts, van der Marel and colleagues \cite{marel03} were able to show that in the low-frequency limit, $\omega < 1.5k_BT/\hbar$, $\tau_{\rm tr} \sim 0.8\tau_P$, in agreement with the dc analysis of Legros \cite{legros19}. 

A second key issue remains, namely; how can such Drude analysis be justified for those strange metals in which the MIR limit is violated and the Drude peak shifts to finite frequencies \cite{husseyMIR}?  Indeed, in the high-$T$ limit, \lq bad metallicity' can be ascribed to a transfer of spectral weight from low- to high-$\omega$, rather than from an ever-increasing scattering rate (that within a Drude picture results in a continuous broadening of the Lorentzian fixed at zero frequency). Given the marked crossover in the form of $\sigma(\omega)$ at low frequencies, it is indeed remarkable and mysterious that the slope of the $T$-linear resistivity continues unabated with no discernible change.

\section{Is it Mottness?}

\begin{figure}
\centering
\includegraphics[scale=0.6]{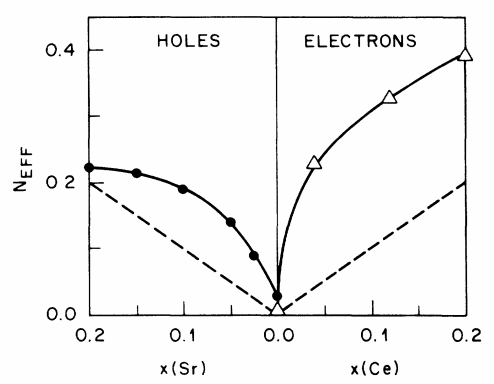}
\caption{Integrated optical conductivity for electron-doped Pr$_{2-x}$Ce${_x}$CuO$_{4-\delta}$ (triangles) and hole-doped La$_{2-x}$Sr$_x$CuO$_{4-\delta}$ (circles) The dashed line
indicates what is expected for a doping a semiconductor.
The expectation is that each Ce or Sr atom contributes just a single charge carrier. Reprinted from Ref.~\cite{CooperUVIR}.}
\label{UVIRmix}
\end{figure}

Table \ref{tab:SMsurvey} encompasses a series of ground states from which $T$-linear resistivity emerges.  In some of these materials, such as the heavy fermions, the high and low-energy features of the spectrum are relatively distinct in the sense that spectral weight transfer from the UV to the IR is absent. On the other hand, hole or electron doping of the parent cuprate induces a marked transfer of spectral weight of roughly 1-2 eV.  As a result, the low-energy spectral weight grows \cite{ctchen,meinders,eskes,PhillipsRMP,CooperUVIR} at the expense of the degrees of freedom at high energy, a trend that persists \cite{marelcolorchange} even inside the superconducting state. This is an intrinsic feature of Mott systems, namely that the number of low-energy degrees of freedom is derived from the high-energy spectral weight. As this physics is distinct from that of a Fermi liquid and intrinsic to Mott physics, it is termed \lq Mottness'\cite{PhillipsRMP}. Notably, the mid-infrared response with its characteristic $\omega^{-2/3}$ scaling is absent from the parent Mott insulating state. Hence, it must reflect the doping-induced spectral weight transfer across the Mott gap. It is perhaps not a surprise then that no low-$T_c$ material exhibits such a significant midinfrared feature. In fact, some theories of cuprate superconductivity \cite{Leggett} credit its origin to the mid-infrared scaling. We can quantify the total number of low-energy degrees of freedom that arise from the UV-IR mixing across the Mott gap by integrating the optical conductivity,
\beq
N_{\rm eff}(\Omega)=\frac{2m V_{\rm cell}}{\pi e^2}\int_0^\Omega
\sigma(\omega)d\omega,
\eeq
up to the optical gap $\Omega\approx$ 1.2 eV where $V_{\rm cell}$ is the unit-cell volume. The energy scale of 1.2 eV corresponds to the minimum of the optical conductivity as mentioned in the previous section.  In a rigid-band semiconductor model in which such spectral weight transfer is absent, $N_{\rm eff}=x$, where $x$ is the number of holes.  In the cuprates, however, $N_{\rm eff}$ exceeds $x$ as shown in Fig. (\ref{UVIRmix}).  This is the defining feature of Mottness \cite{ctchen,meinders,eskes,PhillipsRMP} since it is ubiquitous in Mott systems and strictly absent in weakly correlated metals. Even in many of the strange or quantum critical metals described in Table 1, there is little or no evidence that Mottness is playing any significant role. Such a distinction may thus offer a hint to the source of the uniqueness of the cuprate strange metal. In bad metals, on the other hand, a gradual transfer of low-frequency spectral weight out to energies of order the Mott scale is almost universally observed with increasing temperature \cite{husseyMIR} suggesting that Mottness is one of the key components of bad metallic transport.

The optical response in cuprates tells us that there are degrees of freedom that couple to electromagnetism that have no interpretation in terms of doped holes. That is, they are not local entities as they arise from the mixing of both UV and IR degrees of freedom.  It is such mixing that could account for the lack of any distinctive  energy scale\cite{PhillipsRMP}, that is scale invariance, underlying the strange metal.  Additionally, Lee {\it et al.} showed, also from optical conductivity studies \cite{LeeEM}, that throughout the underdoped regime of the cuprate phase diagram, the effective mass remains constant. As a result, the Mott transition proceeds by a vanishing of the carrier number rather than the mass divergence of the Brinkman-Rice scenario \cite{BRice}. (Note that while quantum oscillation experiments on underdoped cuprates show evidence for mass enhancement \cite{ramshaw15}, this is thought to be tied to the charge order centred around 1/8 doping). Such dynamical mixing between the UV and IR scales in Mott systems is well known to give rise to spectral weight in the lower Hubbard band \cite{meinders,eskes,PhillipsRMP} that exceeds the number of electrons, strictly $1+x$, that the band can hold.  Consequently, part of the electromagnetic response of the strange metal at low energies has no interpretation in terms of electron quasiparticles as it arises strictly from UV-IR mixing. Precisely how such mixing leads to scale-invariant $T-$linear resistivity remains open.

\section{Is it about Gravity?}

\begin{table*}
\begin{ruledtabular}
\caption{Snapshot of current theoretical modeling of the strange metal based on consistency with $T-$ linear resistivity, $\omega^{-2/3}$ scaling of the mid-infrared optical conductivity, quadrature magnetoresistance, extended quantum criticality, and what predictions are made in terms of experimental observables.  The transcription of the abbreviations is as follows: MFL = marginal Fermi liqiud, EFL = ersatz Fermi liqiud, SYK = Sachdev-Ye-Kitaev, AdS/CFT = anti de Sitter space/conformal field theory conjecture, AD/EMD = anomalous dimensions/Einstein-Maxwell-dilaton, HM = Hubbard model, QMC = quantum Monte Carlo, ED = exact diagonalization, CA = cold atoms, DMFT/EDMFT = dynamical mean-field theory/embedded dynamical mean-field theory, A-B = Aharonov-Bohm effect, ECFL = extremely correlated Fermi liquid, and QCP = quantum critical point scenarios.}
\footnote{Apologies to anyone whose work we did not cite.}
\begin{tabular}{ccccccc}
\centering
 & \makecell{$\rho\propto T$} & \makecell{$\rho\propto T$} & \makecell{$\sigma\propto\omega^{-2/3}$} & \makecell{Quadrature} &\makecell{\rm Extended} &\makecell{\rm Experimental} \\
 & \makecell{as $T \rightarrow 0$} & \makecell{as $T \rightarrow \infty$} & \quad & {\rm MR}  &\makecell{criticality}& {\rm Prediction} \\
\colrule
Phenomenological    & & & & & & \\
MFL     & $\checkmark$ \cite{varma96} & $\times$\cite{varma96}             & $\times$             & $\times$  & $\times$  & loop currents \cite{loopvarma}\\
EFL     & - \footnote{$T$-linear resistivity is an input.}                   &                    - &                    - & $\times$  & $\times$  & loop currents \cite{else1}\\
Numerical    & & & & & & \\
ECFL &$\times$ &\checkmark\cite{maishastry} &  -  &  -  &$\times$ &  $\times$\\
HM (QMC/ED/CA)     & -            \cite{Huang987} & $\checkmark$ \cite{Huang987,ED,CA,ED1,ED2} & $\times$      & -                & - & - \\
DMFT/EDMFT    & $\checkmark$ \cite{Cha18341}     & $\checkmark$\cite{kotliarDMFT,tremblay}  & $\times$    & -  & $\checkmark$ \cite{tremblay} & - \\
QCP & \checkmark\cite{INem}&-&-&-&$\times$&- \\
Gravity-based    & & & & & & \\
SYK     & $\checkmark$ \cite{patelPM,syk2} & $\checkmark$\footnote{A slope change occurs through the MIR.} \cite{syk2} & $\times$  & $\checkmark$\footnote{Quadrature scaling obtained only for a bi-valued random resistor model \cite{syk1} with equal weights \cite{boyd19}.} \cite{syk1} & - &$\times$      \\
AdS/CFT & $\checkmark$ \cite{adscftstrange} & $\checkmark$ \cite{adscftstrange} & $\checkmark$\footnote{While this scaling was thought to arise in pure AdS with an inhomogenous charge density \cite{horowitz}, later studies \cite{langley,donos} found otherwise.} \cite{kiritsis,kiritsis2} & $\times$      & $\times$ & $\times$    \\
AD/EMD  & $\checkmark$ \cite{hk,gl1,limtra} & $\checkmark$ \cite{hk,kiritsis,kiritsis2,limtra,karch2} & $\checkmark$ \cite{karch2,kiritsis,kiritsis2}     & $\times$      & $\checkmark$\cite{kiritsis} & Fractional A-B \cite{limtra}
\end{tabular}
\label{SMtheory}
\end{ruledtabular}
\end{table*}
To frame the theoretical modeling of strange metallicity tabulated in Table \ref{SMtheory}, we group the work into three principal categories: 1)  phenomenological, 2) numerical and 3) gravity-related. While both phenomenological models considered here require (EFL) or predict (MFL) loop currents, they do so for fundamentally different reasons. (For an explanation of the various acronyms, please refer to the caption in Table \ref{SMtheory}).  On the EFL account \cite{else1}, such current order is needed to obtain a finite resistivity in the absence of momentum relaxation (certainly not a natural choice given the Drude fit to the optical conductivity discussed previously), while in MFL, loop currents \cite{loopvarma} are thought to underpin the local fluctuation spectrum \cite{varma96}.  ECFL \cite{maishastry} predicts a resistivity that interpolates between Fermi-liquid-like $T^2$ at low $T$ to $T$-linear for $T\gg T_{\rm FL}$.  QMC \cite{Huang987,ED,ED1,ED2} as well as cold atom (CA) experiments \cite{CA} on the Hubbard model (HM) have established that at high temperatures, the resistivity is indeed $T$-linear.  The Fermion-sign problem, however, prevents any definitive statement about the low-$T$ behavior in the presence of Mott physics. Non-Fermi liquid transport in SYK models \cite{SY,Kitaev,K} is achieved by an all-to-all random interaction.  While such interactions might seem initially unphysical, SYK models are nevertheless natural candidates to destroy Fermi liquids which, by their nature, permit a purely local description in momentum space.  As a result, they  are impervious to repulsive local-in-space interactions \cite{polchinski}.  Coupling a Fermi liquid to an array of disordered SYK islands, however, leads \cite{syk1,syk2} to a non-trivial change in the electron Green function across the MIR and hence a change in slope of the resistivity is unavoidable \cite{syk1} though it can be minimized through fine tuning \cite{syk2}.

An added feature of these disordered models is that in certain limits, they have a gravity dual \cite{Kitaev,SY1,K,dual}.  This state of affairs arises because the basic propagator \cite{K,SY1,Kitaev} in the SYK model in imaginary time describes the motion of fermions, with appropriate boundary conditions, between two points of the asymptotic boundary of a hyperbolic plane.  In real time, simply replacing the hyperbolic plane with the space-time equivalent, namely two-dimensional anti de Sitter (AdS) space (a maximally symmetric Lorentzian manifold with constant negative curvature), accurately describes all the correlators.  It is from this realization that the dual description between a random spin model and gravity in AdS$_2$ lies \cite{sachdev,Kitaev,K}. Hence, although the origins of SYK were independent of gravity, its correlators can be deduced from the asymptotics of the corresponding spacetime.  At the asymptote, only the time coordinate survives and hence ultimately, SYK dynamics is ultra-local in space with only diverging correlations in time, an instantiation of local quantum criticality. 

Such local quantum criticality is not a new concept in condensed matter systems and indeed lies at the heart of MFL phenomenology \cite{varma96}, DMFT \cite{kotliarDMFT}, and is consistent with the momentum-independent continuum found in the M-EELS data discussed earlier \cite{mitrano18}. The deeper question is why does gravity have anything to do with a spin problem with non-local interactions?  The issue comes down to criticality and to the structure of general relativity. The second equivalence principle on which general relativity is based states that no local measurement can detect a uniform gravitational field.  A global measurement is  required.  Ditto for a critical system since no local measurement can discern criticality.  Observables tied to the diverging correlation length are required. Hence, at least conceptually, it is not unreasonable to expect a link between critical matter and gravity. The modern mathematical machinery which makes it possible to relate the two is the gauge-gravity duality or the AdS/CFT (conformal field theory) conjecture. The key claim of this duality \cite{maldacena,witten,gubser} is that some strongly interacting quantum theories, namely ones which are at least conformally invariant in $d$-dimensions, are dual to a theory of gravity in a $d+1$ spacetime that is asymptotically AdS. The radial direction represents the energy with the quantum theory residing at the UV boundary and the IR limit deep in the interior at the black hole horizon.  Hence, intrinsic to this construction is a separation between bulk (gravitational) and boundary (quantum mechanical) degrees of freedom. That the boundary of a gravitational object has features distinct from the bulk dates back to the observations of Beckenstein \cite{beckenstein} and Hawking \cite{hawking,hawkingarea} that the information content of a black hole scales with the area, not the volume.  The requirement that the boundary theory be strongly coupled then arises by maintaining that the AdS radius exceeds the Planck length $\ell_P$.  More explicitly, because the AdS radius and the coupling constant of the boundary theory are proportional, the requirement $R\gg\ell_P$ translates into a boundary theory that is strongly coupled. 

The first incarnation \cite{Faulkner,fireball,schalm} of this duality in the context of fermion correlators involved modeling fermions at finite density in $2+1$ dimensions.  From the duality, the  conformally invariant vacuum of such a system corresponds to gravity in AdS$_4$, the extra dimension representing the radial direction along which identical copies of the boundary CFT lie albeit with differing energy scales.  Surprisingly, what was shown \cite{Faulkner} is that the low-energy (IR) properties of such a system in the presence of a charge density are determined by an emergent AdS$_2\times R^2$ (with $R^2$ representing a plane) spacetime at the black hole horizon.  The actual symmetry includes scale invariance and is denoted by $SL(2,R)$ (a special Lie group of real $2\times 2$ matrices with a unit determinant). Once again, the criticality of boundary fermions is determined entirely by the fluctuations in time, that is, local quantum criticality as seen in SYK.  The temperature and frequency dependence of the conductivity are then determined by the same exponent \cite{Faulkner} as expected from Eqs. (\ref{qctemp}) and (\ref{qcomega}) and as a result, a simultaneous description of $T$-linearity and $\omega^{-2/3}$ is not possible, as noted in Table \ref{SMtheory}. 

This particular hurdle is overcome by the AD/EMD theories \cite{Anomdim0,Anomdim01,Anomdim02,kiritsis,kiritsis1,kiritsis2,cremonini,Anomdim1} which as indicated in Table \ref{SMtheory}, have been the most successful to date in describing the range of physics observed in strange metals.  What is new here is the introduction of extra fields, dilatons for example, which permit hyperscaling violation\cite{shl} and anomalous dimensions\cite{Anomdim0,Anomdim01,Anomdim02,kiritsis,kiritsis1,kiritsis2,cremonini,Anomdim1} for all operators.  Consequently, under a scale change of the coordinates, the metric is no longer unscathed. That is, the manifold is not fixed and it is the matter fields that determine the geometry.  Such systems have a covariance, rather than scale invariance indicative of pure AdS metrics.  A consequence of this covariance is that even the currents acquire anomalous dimensions.  But how is this possible given that a tenet of field theory is that no amount of renormalization can change the dimension of the current \cite{gross} from $d-1$?  What makes this possible is that in EMD theories, the extra radial dimension allows physics beyond Maxwellian electro-magnetism.  For example, the standard Maxwell action, $S=\int dV_d F^2$ where $F=dA$,  requires that the dimension of the gauge field be fixed to unity, $[A]=1$\footnote{What is really required is that $[qA]=1$, with $q$ the charge.  In insisting that $[A]=1$, we are setting $q=1$ but still all of our statements refer to the product $qA$.}. EMD theories use instead an action of the form $S=\int dV_d dy y^a F^2$ where $y$ is the radial coordinate of the $d+1$ AdS spacetime. Comparing these two actions leads immediately to the conclusion that the dimension of $A$ now acquires the value $[A]=1-a/2$. Hence, even in the bulk of the geometry, the dimension of the  gauge field is not unity.  Depending on the value of $a$, $a<0$ at the UV conformal boundary or $a>0$ at the IR at the black hole horizon, the equations of motion are non-standard and obey fractional electromagnetism \cite{gl1,gl2} consistent with a non-traditional dimension for the gauge field. In EMD theories, it is precisely the anomalous dimension\cite{kiritsis,kiritsis1,kiritsis2,cremonini,Anomdim0,Anomdim01,Anomdim02} for conserved quantities that gives rise to the added freedom for extended quantum criticality to occur, the simultaneous fitting \cite{karch2} of $T-$linearity and $\omega^{-2/3}$ of the optical conductivity, and the basis for a proposal for the strange metal based on $[A]=5/3$ \cite{hk}. 

Within these holographic systems, a Drude-like peak in the optical conductivity can emerge both from the coherent (quasiparticle-like) sector \cite{Davison_15} as well as the incoherent (\lq un-particle \cite{unparticle}') sector \cite{hartnoll10, kiritsis15, chen_17, Davison_19}.  Application of EMD theory has also provided fresh insights into the phenomenon of \lq lifetime separation' seen in the dc and Hall conductivities of hole-doped cuprates \cite{carrington92, chien91, manako92} as well as in other candidate strange metals \cite{paschen04,lyu20}. For a system with broken translational invariance, the finite density conductivity comprises two distinct components \cite{blake_donos}, with the dc resistivity being dominated by the particle-hole symmetric term -- with a vanishing Hall conductivity -- and one from an explicit charge density governed by more conventional (Umklapp) momentum relaxation that sets the $T$-dependence of the Hall angle.  

The success of EMD theories in the context of strange metal physics raises a philosophical question:  Is all of this just a game? That is, is the construction of bulk theories with funky electromagnetism fundamental? The answer here lies in N\"other's Second Theorem (NST) \cite{gl1,gl2,PhillipsRMP}, a theorem far less known than her ubiquitous first theorem but ultimately of more importance as it identifies a shortcoming.  To illustrate her first theorem, consider  Maxwellian electromagnetism which is invariant under the transformation $ A_\mu\rightarrow A_\mu+\partial_\mu\Lambda$. This theorem states that there must be a conservation law with the same number of derivatives as in the gauge principle. Hence the conservation law only involves  a single derivative, namely $\partial_\mu J_\mu=0$. This is N\"other's First Theorem \cite{N} in practice. 

What N\"other \cite{N} spent the second half of her famous paper trying to rectify is that the form of the gauge transformation is not unique; hence the conservation law is arbitrary.  It is for this reason that in the second half \cite{N} of her foundational paper, she retained all possible higher-order integer derivatives in the gauge principle.  These higher-order derivatives both add constraints to and change the dimension of the current.  Stated  succinctly, NST \cite{N} dictates that the full family of generators of U(1) invariance determines the dimension of the current.  It is easy to see how this works.  Suppose we can find a quantity, $\hat Y$ that commutes with $\partial_\mu$.  That is, $\partial_\mu \hat Y=\hat Y\partial_\mu$.  If this is so, then we can insert this into the conservation law with impunity.  What this does is redefine the current: $\partial_\mu \hat Y J^\mu=\partial_\mu \tilde J^\mu$.  The new current $\tilde J^\mu$ acquires whatever dimensions $\hat Y$ has such that $[\tilde J^\mu]=d-1-d_Y$.  But because of the first theorem, $\hat Y$ must have come from the gauge transformation and hence must ultimately be a differential operator itself.  That is, there is an equally valid class of electromagnetisms with gauge transformations of the form $A_\mu\rightarrow A_\mu+\partial_\mu\hat Y\Lambda$. For EMD theories \cite{gl1,gl2,PhillipsRMP}, $\hat Y$ is given by the fractional Laplacian, $\Delta^{(\gamma-1)/2}$ with $[A_\mu]=\gamma$ (with $\gamma=1-a/2$ to make contact with the EMD theories introduced earlier).  For most matter as we know it, $\gamma=1$. The success of EMD theories raises the possibility that the strangeness of the strange metal hinges on the fact that $\gamma\ne 1$. This can be tested experimentally using the standard Aharonov-Bohm geometry \cite{limtra,gl1} in which a hole of radius $r$ is punched into a cuprate strange metal.  Because $[A]$ is no longer unity, the integral of $A\cdot d\ell$ is no longer the dimensionless flux.  For physically realizable gauges, this ultimately provides an obstruction to charge quantization.  As a result, deviations \cite{limtra,gl1} from the standard $\pi r^2\times B$ dependence for the flux would be the key experimental feature that a non-local gauge principle is operative in the strange metal. An alternative would be, as Anderson \cite{anderson} advocated, the use of fractional or unparticle propagators with the standard gauge principle. However, in the end, it all comes down to gauge invariance.  The standard gauge-invariant condition prevents the power laws in unparticle stuff from influencing the algebraic fall-off of the optical conductivity \cite{limtragool,karch2} as they offer just a prefactor to the polarizations \cite{Liao2008}.  The escape route, an anomalous dimension for the underlying gauge field, offers a viable solution but the price is abandoning locality \cite{bora} of the action. \\

\begin{figure}
\centering
\includegraphics[scale=0.3]{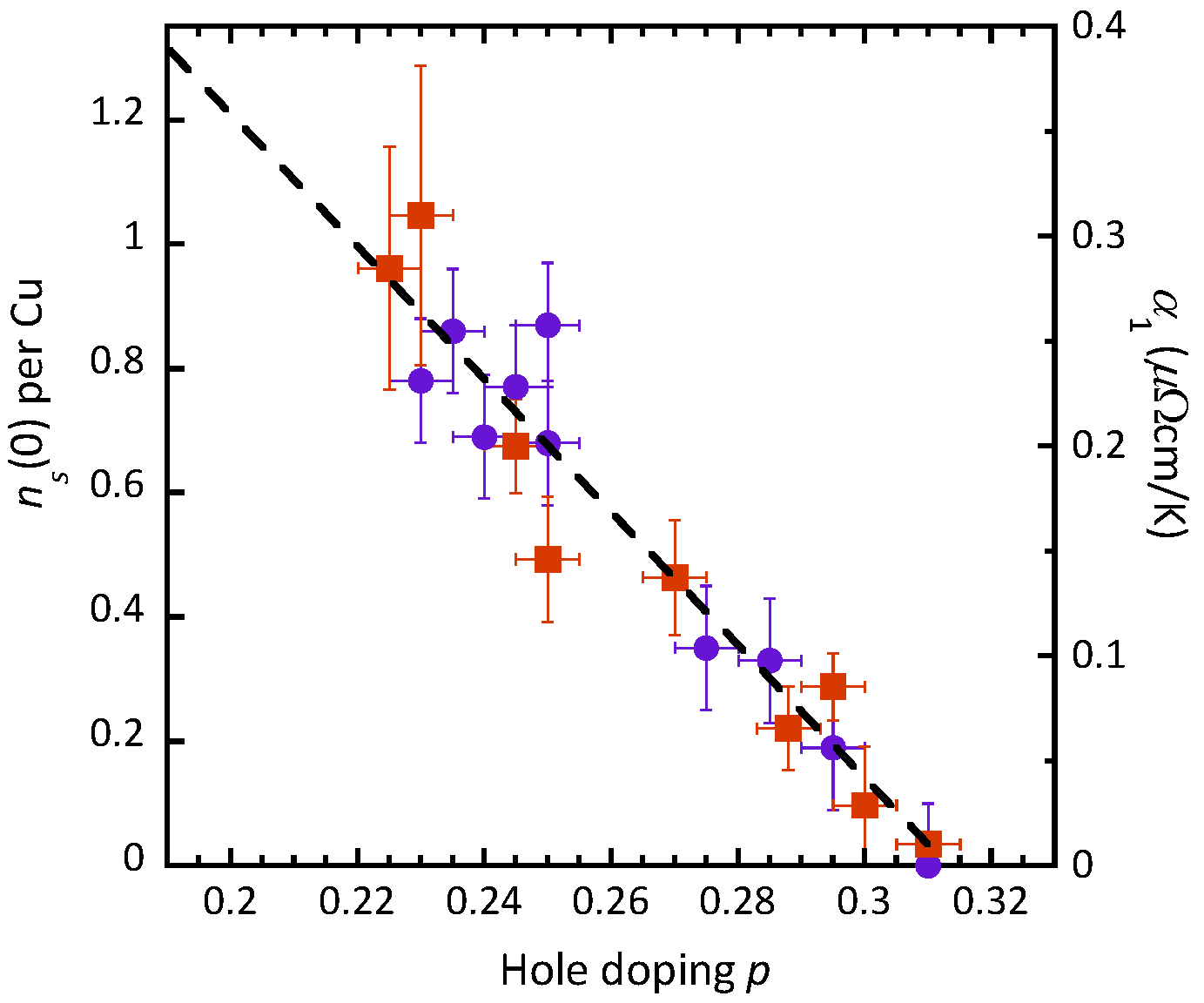}
\caption{
Correlation between the superfluid density $n_s(0)$ and the coefficient $\alpha_1$ of the $T$-linear resistivity in Tl$_2$Ba$_2$CuO$_{6+\delta}$ (Tl2201) across the strange metal regime (adapted from Refs.~\cite{culo21, putzke21}).}
\label{tl2201}
\end{figure}

\section{Is it Important?}

Given the immense difficulty in constructing a theory of the strange metal, one might ask why bother?
To gauge the importance of the strange metal, look no further than Fig. (\ref{tl2201}).  This figure shows that the coefficient $\alpha_1$ of the $T$-linear resistivity component in the strange metal regime of overdoped hole-doped cuprates tracks the doping dependence of the $T=0$ superfluid density $n_s(0)$. As mentioned earlier, a similar correlation exists between $\alpha_1$ and $T_c$ in electron-doped cuprates \cite{jin11}, the Bechgaard salts \cite{doiron09} as well as the iron pnictides \cite{doiron09}, establishing a fundamental link between high-temperature superconductivity and the strange metal.

For a long time, the drop in $n_s(0)$ with doping in cuprates was attributed to pair breaking, a symptom of the demise of the dominant pairing interaction within a disordered lattice. Recent mutual inductance measurements, however, have challenged this view, arguing that the limiting low-$T$ behavior of $n_s(T)$ was incompatible with conventional pair breaking scenarios \cite{bozovic16}. Certainly, the correlation between $\alpha_1$ and $n_s(0)$ is unforeseen in such models. Moreover, if the strange metal regime is indeed populated with non-quasiparticle states, then Fig. (\ref{tl2201}) indicates a pivotal role for these states in the pairing condensate \cite{culo21}. On more general grounds, this result informs us that the door to unlocking cuprate superconductivity is through the strange metal and any theory which divorces superconductivity from the strange metal is a non-starter.  To conclude, solving the strange metal kills two birds with one stone.  Perhaps there is some justice here.  After all, we know from Pippard's \cite{pippard} work, which can be reformulated \cite{gl1,gl2} in terms of fractional Laplacians, that even explaining superconductivity in elemental metals necessitates a non-local relationship between the current and the gauge field. What seems to be potentially new about the cuprates is that now the normal state as a result of the strange metal also requires non-locality. 

\textbf{Acknowledgements} 

 PA and PWP acknowledge support from Center for Emergent Superconductivity, a DOE Energy Frontier Research Center, Grant No. DE-AC0298CH1088. NEH is funded by Netherlands Organisation for Scientific Research (NWO) (\lq Strange Metals' 16METL01), the European Research Council (ERC) under the European Union’s Horizon 2020 research and innovation programme (No. 835279-Catch-22) and EPSRC (EP/V02986X/1). The work on fractional electromagnetism was funded through DMR-2111379. 
\bibliographystyle{Science}
\bibliography{strange}
\end{document}